\documentclass[12pt]{article}
\usepackage{latexsym}
\usepackage{amssymb}
\usepackage{amsfonts}
\newcommand{\be}{\begin{equation}}
\newcommand{\ee}{\end{equation}}
\newcommand{\bea}{\begin{eqnarray}}
\newcommand{\eea}{\end{eqnarray}}

\newcommand{\gm}{\gamma}
\newcommand{\Gm}{\Gamma}

\newcommand{\ep}{\varepsilon}
\newcommand{\sg}{\sigma}

\newcommand{\lm}{\lambda}

\newcommand{\dd}{\mbox{d}}

\newcommand{\nn}{\nonumber}

\begin{document}
\parindent=1.5pc

\begin{titlepage}

\bigskip
\begin{center}
{{\large\bf
On the Resolution of Singularities of Multiple \\
Mellin--Barnes Integrals
} \\
\vglue 5pt
\vglue 1.0cm
{\large  A.V. Smirnov}\footnote{E-mail: asmirnov80@gmail.com}\\
\baselineskip=14pt
\vspace{2mm}
{\normalsize Scientific Research
Computing Center of Moscow State University
   }\\
\baselineskip=14pt
\vspace{2mm}
and\\
\baselineskip=14pt
\vspace{2mm}
{\large   V.A. Smirnov}\footnote{E-mail: smirnov@theory.sinp.msu.ru}\\
\baselineskip=14pt
\vspace{2mm}
{\normalsize
Nuclear Physics Institute of Moscow State University\\
}
\baselineskip=14pt
\vspace{2mm}
\vglue 0.8cm
{Abstract}}
\end{center}
\vglue 0.3cm
{\rightskip=3pc
 \leftskip=3pc
\noindent
One of the two existing strategies of resolving singularities of
multifold Mellin--Barnes integrals in the dimensional
regularization parameter, 
or a parameter of the analytic
regularization, is formulated in a modified
form. The corresponding algorithm is implemented as a {\tt Mathematica}
code {\tt MBresolve.m}
\vglue 0.8cm}
\end{titlepage}

\section{Introduction}

The method of Mellin--Barnes (MB) representation is one of the most powerful
methods to evaluate multiloop Feynman integrals.
It is based on a very simple formula,
\bea
\frac{1}{(X+Y)^{\lm}} = \frac{1}{\Gm(\lm)}
\frac{1}{2\pi i}\int_{-i \infty}^{+i \infty} \dd z\,
\Gm(\lm+z) \Gm(-z) \frac{Y^z}{X^{\lm+z}}  \; ,
\label{MB}
\eea
which is called MB representation and is applied to replace a sum of two terms raised
to some power by their products in some powers.
The contour of integration should be chosen so that
the poles with a $\Gm(\ldots+z)$ dependence
are to the left of the contour and
the poles with a $\Gm(\ldots-z)$ dependence
are to the right of it.

The simplest way to apply (\ref{MB}) is to represent massive
propagators as continuous superpositions  of massless ones, at the cost of
introducing additional integrations over MB variables. More often,
MB integrals are introduced at the level of Feynman or alpha
parameters. Anyway, (\ref{MB}) is used in an appropriate way, with the goal to obtain
integrals over loop momenta or Feynman/alpha parameters which can
be taken in terms of gamma functions. As a result one obtains a
multiple MB integral with an integrand involving gamma functions
and powers of kinematic invariants.

It is important to derive a MB representation for general values
of the powers of the propagators. First, this provides unambiguous
prescriptions for a choice of contours of integration. The
resulting rule is similar to the rule for (\ref{MB}): when we
understand a multiple MB integral iteratively and consider an
integration over a variable $z_i$ it is implied that the poles
with a $\Gm(\ldots+z_i)$ dependence are to the left of the contour
and the poles with a $\Gm(\ldots-z_i)$ dependence are to the right
of it. Second, such a general MB representation provides crucial
checks and can be applied at any chosen values of the powers of
the propagators.

When evaluating a Feynman integral for specific powers of
the propagators (indices), one starts from the general MB representation and
obtains an integral of the form
\bea
\frac{1}{(2\pi i)^n} \int_{-i\infty}^{+i\infty}\ldots \int_{-i\infty}^{+i\infty}
\frac{\prod_i \Gm\left(a_i+b_i \ep+\sum_j c_{ij} z_j\right)}
{\prod_i \Gm\left(a'_i+b'_i \ep+\sum_j c'_{ij} z_j\right)}
\prod_k x_k^{d_k} \prod_{l=1}^n \dd z_l
\label{MBn}
\;,
\eea
where $\ep=(4-d)/2$ is the dimensional regularization parameter,
$a_i,\ldots,c'_{ij}$ are rational numbers, $x_k$ are
ratios of kinematic invariants and/or masses, and their exponents,
$d_k$, are linear combinations of $\ep$ and $z$-variables.
Typically, $c_{ij}=\pm 1$.

Although results for dimensionally regularized Feynman integrals
are practically needed in a Laurent expansion in $\ep$, one could
try to evaluate a given MB integral (\ref{MBn}) for general $\ep$.
Then the $\ep$-expansion is performed in the result. However, for
sufficiently complicated multiloop Feynman integrals, the
evaluation at general $\ep$ turns out to be impossible, and one
proceeds with an $\ep$-expansion.

One of the advantages of this method is that the
singularity structure in $\ep$ can be resolved in a simple way.
This procedure basically consists of taking residues and shifting
contours, with the goal to obtain a sum of integrals where one can
expand integrands in Laurent series in $\ep$.
To do this one can apply two
strategies formulated in \cite{MB1} and \cite{MB2} which we are going
to call {\em Strategy~A} and {\em Strategy~B}, respectively.

Strategy~A is described and illustrated in numerous examples in Chapter~4 of
\cite{books}. It was applied, e.g., in \cite{SAappl}.
According to strategy~A, one performs an analysis of the
integrand to reveal how poles in $\ep$ arise. The guiding
principle is that the product $\Gm(a+z)\Gm(b-z)$,
where $a$ and $b$ can depend on the rest of the integration
variables, generates, due to
the integration over $z$, the singularity of the type
$\Gm(a+b)$.
Indeed, if we shift an initial contour of integration over $z$
across the point $z=-a$ we obtain an integral over a new contour
which is not singular at $a+b=0$, while the corresponding residue
involves an explicit factor $\Gm(a+b)$.
This observation shows that any contour of one of
the following integrations over the rest of the MB variables should be
chosen according to this dependence, $\Gm(a+b)$.
Hence one thinks of integrations in various orders and
then identifies some `key' gamma functions which are crucial
for the generation of poles in $\ep$. Then one
takes residues and shifts contours, starting from the first poles of
these key gamma functions.
The same analysis and procedure is applied to the contributions of the
residues.

Within Strategy~B, one chooses an initial value
of $\ep$ and values of
the real parts of the integration variables, $z_i,\ldots$ in
such a way that the real parts of all the arguments of the gamma
functions in the numerator are positive and
one can integrate over straight lines.
Then one tends $\ep$ to zero and whenever
the real part of the argument
of some gamma function vanishes one crosses this pole and
adds a corresponding residue which has one integration less and
is treated as the initial integral within the same procedure.

Strategy~B  is algorithmic in its character and, indeed,
two algorithmic descriptions were formulated in \cite{AnDa,Czakon}.
A public code called {\tt MB.m} was presented in \cite{Czakon}.
Strategy~B was successfully applied, e.g., in \cite{SBappl,CMM} and in many other
papers.

The purpose of this letter is to formulate Strategy~A in a slightly
modified form and to present the corresponding algorithm implemented in
{\tt Mathematica}.

\section{The modified Strategy~A}

To present a modified Strategy~A let us explicitly formulate what was
implied in the
initial Strategy~A. When we take care of one of the key gamma
functions we shift a contour and take a residue. Let $\Gm(A_i)$
with $A_i= a_i+b_i \ep+\sum_j c_{ij} z_j$ be one of the key gamma
functions in (\ref{MBn}). Without loss of generality, we can
consider $\ep$ real. Then changing the nature of the first pole of
this gamma function means changing the rule for an admissible
contour, i.e. that, instead of the condition Re$A_i>0$ when
crossing the real axis in the process of the integration, we have
the condition  $-1<\mbox{Re}\,A_i<0$. Let us denote this
transition by replacing $\Gm(A_i)$ by $\Gm^{(1)}(A_i)$. The initial rule for
the contour can be changed again and then we have the condition
$-n<\mbox{Re}\,A_i<-n+1$ for $n=2,3,\ldots$ with the notation
$\Gm^{(n)}(A_i)$.

Within Strategy~B one has straight contours in the beginning.
Rather, in the modified Strategy~A, we will be oriented at
straight contours in the end. Apparently,  it is desirable to
achieve a minimal number of terms after the resolution of the
singularities in $\ep$. To do this, let us try to search for
contours which are going to have in the end of this procedure and for which
the gamma functions in the numerator are changed, in the above
sense, in a minimal way.

To formalize this requirement, let us introduce the function
$\sg(x)=[(1-x)_+]$ where $[\ldots]$ is the integer part of a number
and $x_+=x$ for $x>0$ and $0$, otherwise. In other words, if
$-n<x<-n+1$ then $\sg(x)=n$ for $n>0$ and $\sg(x)=0$ for $n\leq 0$.

So, let us set $\ep=0$ and search for contours, i.e. Re$z_i$, for
which the sum
\[\sum_i \sg\left(\left. \mbox{Re}A_i\right|_{\ep=0}\right)\equiv
\sum_i \sg\left( a_i+\sum_j c_{ij} \mbox{Re}z_j\right)\]
is minimal.

After such a choice is done we identify gamma functions 
which should be
changed, in the above sense, in order to arrive at a final
integral where a Laurent expansion in $\ep$ is possible. In fact,
this step replaces the first step in the primary Strategy~A where
one identified such key gamma functions after the analysis
characterized above.

Then the second step in Strategy~A is the same as in the old
version:
we take care of the distinguished gamma functions, i.e. take a residue
and replace $\Gm$ by  $\Gm^{(1)}(A_i)$ (and, possibly,
$\Gm^{(1)}(A_i)$ by  $\Gm^{(2)}(A_i)$ etc.) We proceed
iteratively, as in the previous strategy: every residue is
considered from scratch, i.e. treated in the same way as the initial MB
integral.

Let us emphasize that although this strategy aims to minimize the number of
resulting terms, we cannot exclude that there is another way of resolving the
singularities in $\ep$ that is the best one in this sense. For example, it can
happen, in rather complicated examples, that different orders of changing the
key gamma functions lead to different numbers of resulting terms. Still we
believe that such a difference is negligible and that the Strategy~A provides a
resolution of the singularities in $\ep$ at least very close to the theoretically
best one.

The difference of the new and the old Strategies~A is minor. In the
examples of \cite{SAappl} one can see that resulting contours
were straight indeed. This difference can still be seen in the
following simple example of the integral
\bea
\frac{1}{2\pi i} \int_{-i\infty}^{+i\infty}
\Gm(1+\ep+z) \Gm(-1/2 + \ep + z) \Gm(3/2 -\ep-z) \Gm(-z)
\; \dd z
\label{ExA}
\nn
\eea
which, of course, can be evaluated at general $\ep$ by the first
Barnes lemma. Within the `old' Strategy~A, one can observe that
there is no gluing of poles so that one can expand the integrand
in $\ep$. However, a resulting contour cannot be chosen as a straight
line. Rather, within the `new' Strategy~A, we have to choose a
gamma function to be modified and, as a result, we obtain a
residue and an integral over a straight line which both can be
expanded in $\ep$.

\section{Implementation in {\tt Mathematica}}

The code should be loaded together with the package {\tt MB} \cite{Czakon} by
Czakon. It uses the {\tt MBresidue} routine and some other private
functions from that package; after obtaining the list of integrals one can
continue the contour optimization and the numerical evaluation with the
standard routines of {\tt MB}.

Formally the code can be described the following way: the initial function
results in contour prescriptions (the arguments of $\Gamma$-functions have
to be positive in the end). Now

{\tt MBresolve}$[${\tt function},{\tt cont\_pr}$]$
\begin{tabbing}
1. \= find a point with the minimal number of contour prescriptions broken.\\
2. \> \textbf{If} this number is equal to zero, \textbf{then return} {\tt function} as an answer\\
3. \> $l =$ list of broken contour prescriptions\\
4. \> \textbf{For} \= each $\{x,n\}$ in $l$ do \\
5. \> \> \textbf{Try} \= to evaluate {\tt t1 = }{\tt MBresolve}$[${\tt function},{\tt cont\_pr}$']$ \\
 \> \> \>  where {\tt cont\_pr}$'$ are obtained by replacing $\{x,n\}$ with $\{x,n+1\}$ \\
6. \> \> \textbf{On exception} move to the next For cycle \\
7. \> \> represent $x$ as $c\alpha + r$, where $c\neq 0$, $\alpha$ is one of the integration variables \\
   \> \> and $r$ does not depend on $\alpha$ \\
8. \> \> \textbf{Return} \= {\tt t1 - Sign}$[c]${\tt MBresolve}$[${\tt MBresidue}$[${\tt function}$,\{\alpha,-r/c\}],${\tt cont\_pr}$'']$, \\
   \> \> \> where {\tt cont\_pr}$''$ are inherited from {\tt cont\_pr} by removing $\{x,n\}$ \\
   \> \> \> and performing the substitution $\alpha\rightarrow -r/c$ in all other functions \\
9. \> \textbf{End} \\
10. \> \textbf{Throw exception}
\end{tabbing}

The {\tt function} should be a function of integration variables such that the poles
are only due to the {\tt Gamma} and {\tt PolyGamma}  factors
(with linear arguments). The contour prescriptions {\tt cont\_pr} are a list of
pairs ${x,n}$ where $x$ is a linear function of integration variables and $n$ is an
non-negative integer. If $n$ is equal to zero, such a term means that $x$ has to be positive;
if $n$ is positive, then $x$ has to be greater that $-n$ and smaller than $-n+1$.
If the {\tt cont\_pr} parameter is missing, it is created automatically
by considering all arguments of {\tt Gamma} functions and their derivatives
and assuming them all to be positive.
An uncaught exception at the top level appears if there is a degenerate case
and one is required to introduce extra regularization parameters.

Hence, instead of using {\tt MBoptimizedRules[]} or {\tt MBrules[]} and,
subsequently, {\tt MBcontinue[]}, within {\tt MB.m} \cite{Czakon}, one
can now apply {\tt MBresolve[]}.

The  corresponding {\tt Mathematica} code {\tt MBresolve.m}
is public and can be found at
http://projects.hepforge.org/mbtools/ together with other tools for evaluating
MB integrals.


Here is an example of a tenfold MB representation derived loop by
loop for the four-loop ladder massless on-shell diagram with $p_i^2=0, \;
i=1,2,3,4$, where $s=(p_1+p_2)^2$  and $t=(p_1+p_3)^2$.
\begin{verbatim}
In[1]:
<< MB.m;
<< MBresolve.m;
F = -(((-s)^(-5 - 4*ep - z7)*(-t)^z7*Gamma[1 + z1]*
      Gamma[-1 - ep - z1 - z2]*Gamma[-z2]*Gamma[-1 - ep - z1 - z3]*
      Gamma[-z3]*Gamma[1 + z1 + z2 + z3]*Gamma[2 + ep + z1 + z2 + z3]*
      Gamma[z10 - z4]*Gamma[-z1 + z4]*Gamma[-ep + z1 + z2 - z4 - z5]*
      Gamma[-z5]*Gamma[-ep + z1 + z3 - z4 - z6]*Gamma[-z6]*
      Gamma[1 + z4 + z5 + z6]*
      Gamma[1 + ep - z1 - z2 - z3 + z4 + z5 + z6]*Gamma[-z7]*
      Gamma[1 + z7]*Gamma[-z10 + z7]*Gamma[-ep - z10 + z4 + z5 - z8]*
      Gamma[-z8]*Gamma[-ep + z10 - z7 + z8]*
      Gamma[-ep - z10 + z4 + z6 - z9]*
      Gamma[1 + ep - z10 + z7 - z8 - z9]*Gamma[-z9]*
      Gamma[-ep + z10 - z7 + z9]*Gamma[1 + z10 + z8 + z9]*
      Gamma[1 + ep + z10 - z4 - z5 - z6 + z8 + z9])/(Gamma[-2*ep]*
      Gamma[1 - z2]*Gamma[1 - z3]*Gamma[1 - 2*ep + z1 + z2 + z3]*
      Gamma[1 - z5]*Gamma[1 - z6]*Gamma[1 - 2*ep + z4 + z5 + z6]*
      Gamma[1 - z8]*Gamma[1 - z9]*Gamma[1 - 2*ep + z10 + z8 + z9]));
Fcont = MBresolve[F, ep];
Length[Fcont]

MB 1.2
by Michal Czakon
improvements by Alexander Smirnov
\end{verbatim}
\vspace{-0.31cm}
{\tt more info in hep-ph/0511200}
\vspace{-0.31cm}
\begin{verbatim}
last modified 2 Jan 09
MBresolve 1.0
by Alexander Smirnov
\end{verbatim}
\vspace{-0.31cm}
{\tt more info in arXiv:0901.0386}
\vspace{-0.31cm}
\begin{verbatim}
last modified 4 Jan 09
CREATING RESIDUES LIST..........653.5156 seconds
EVALUATING RESIDUES..........15.2969 seconds

Out[1]: 656
\end{verbatim}

\section{Discussion and perspectives}

Let us remind that, sometimes, a given MB representation can be
ill-defined even for a well-defined Feynman integral, in the sense
that a gluing of poles is present, i.e. one can distinguish a
subproduct of the gamma functions such that the sum of there
arguments, at general $\ep$, is equal to a non-positive integer number.
To cure such a MB representation, one can introduce an auxiliary
analytic regularization into some index, i.e. $a_i\to a_i+y$ and then analytically
continue the given MB integral, first, in $y$ to the point $y=0$
and then $\ep$ to the point $\ep=0$. Let us stress that in such
situation one starts, within the new Strategy~A, with setting
$y=0,\ep=0$ and then searches for appropriate contours for which the
gamma functions in the numerators are changed in the minimal way,
similarly to the case without such analytic regularization.

Let us emphasize that both Strategy~A and Strategy~B are based on the
fact that singularities in $\ep$ can be generated, because of
gluing of poles of different nature, by the integration at {\em
compact} regions. At least for planar diagrams, for which one can
apply the code called {\tt AMBRE} \cite{AMBRE} based on the
loop-by-loop strategy, this looks to be the only source of the
singularities. However, for MB representations derived within the
loop-by-loop strategy for nonplanar diagrams, there can be another
source of poles.\footnote{See also a similar discussion in \cite{CMM}.}
This feature can be exemplified by the massless
two-loop nonplanar diagram with two external legs
on-shell.\footnote{It was first calculated in an expansion in $\ep$
up to $\ep^0$ in
\cite{Gons}. Now, results of expansion up to $\ep^2$ \cite{MMV},
up to $\ep^4$ \cite{2lffgenEp} and even a result
for general $\ep$ \cite{2lffgenEp} are available.}
In the corresponding
MB representation derived within the loop-by-loop strategy one meets, in
particular, the following onefold MB integral
\bea
\frac{1}{2\pi i} \int_{-i\infty}^{+i\infty}
 \frac{\Gm(1 + 2\ep + z)\Gm(-z)}{1 + \ep + z}
\; e^{-i\pi z} \dd z \;.
\label{MBnp}
\nn
\eea
There is no gluing of poles so that a pole in $\ep$ cannot be
generated by the integration over finite regions. Still a pole is
generated and this can be seen by an explicit evaluation of this
integral by closing the integration contour to the right and
summing up the resulting series. This can be seen also by analyzing
the asymptotic behaviour of the integrand at infinity. Setting
$z=x+i y$ and using formulae of the asymptotic behaviour of gamma
functions at large arguments in the complex plane, we can observe
that the leading asymptotic behaviour when $y \to +\infty$ is
$1/y^{1-2\ep}$ which explains the appearance of the pole.
Let us still mention that, for this concrete diagram, there is a
better way to obtain a `good' MB representation --- see, e.g.,
Chapter~4 of \cite{books}.

For simple MB integrals, say, up to six-fold, the codes {\tt MB.m}
and {\tt MBresolve.m} work more or less in the same way. For
higher-fold MB integrals, e.g. ten-fold ones, {\tt MBresolve.m} produces at first
fewer integrals after the
resolution of singularities in $\ep$. However, the command
{\tt MBmerge} of {\tt MB.m} combines many terms, and the number of
resulting integrals becomes smaller that produced by {\tt MBresolve.m}.
We should bear in mind, though, that in the current version of {\tt MB.m},
this command combines all integrals which have the same integration
contours, while it is more natural to consider
separately integrals with different patterns of gamma functions,
both from analytical and numerical points of view.
Presumably, the {\tt MBmerge} command can be improved in this respect.

{\em Acknowledgments.}
We are grateful to  M.~Czakon and D.~Kosower for helpful discussions
and to M.~Czakon for careful reading of a draft version of the paper.
Many thanks to N.~Glover for kind hospitality during our visit to the
Institute of Particle Physics Phenomenology in Durham, where a
part of this work was done.
The work was supported by the Russian Foundation for Basic
Research through grant 08-02-01451.


\begin{thebibliography}{99}
\bibitem{MB1}
V.A.~Smirnov,
{\em Analytical result for dimensionally regularized massless on-shell  double
box,  Phys. Lett.} {\bf  B  460} (1999) 397
[arXiv:hep-ph/9905323].

\bibitem{MB2}
J.B.~Tausk,
{\em Non-planar massless two-loop Feynman diagrams with four on-shell legs,
Phys. Lett.} {\bf B 469} (1999) 225
[arXiv:hep-ph/9909506].

\bibitem{books}
V.A.~Smirnov, {\em Evaluating Feynman Integrals},
Springer Tracts Mod.\ Phys.\  {\bf 211} (2004) 1;
V.A.~Smirnov, {\em Feynman integral calculus},
(Berlin, Germany, Springer, 2006).

\bibitem{SAappl}
V.A.~Smirnov and O.L.~Veretin,
{\em Analytical results for dimensionally regularized massless on-shell  double
boxes with arbitrary indices and numerators,
Nucl. Phys.}  {\bf B 566} (2000) 469
[arXiv:hep-ph/9907385];
\\
V.A.~Smirnov,
{\em Analytical Result for Dimensionally Regularized Massless Master Double Box
with One Leg off Shell,
Phys. Lett.}  {\bf B 491} (2000) 130
[arXiv:hep-ph/0007032];
{\em Analytical result for dimensionally regularized massless master  non-planar
double box with one leg off shell,
Phys. Lett.} {\bf B 500}, (2001) 330
[arXiv:hep-ph/0011056];
{\em The leading power Regge asymptotic behaviour of dimensionally regularized
massless on-shell planar triple box,
Phys. Lett.} {\bf B 547} (2002) 239
[arXiv:hep-ph/0209193];
{\em Analytical result for dimensionally regularized massless on-shell  planar
triple box,
Phys. Lett.} {\bf B 567} (2003) 193
[arXiv:hep-ph/0305142];
{\em Analytical result for dimensionally regularized massive on-shell planar
double box,
Phys. Lett.} {\bf B 524} (2002) 129
[arXiv:hep-ph/0111160];
{\em Evaluating multiloop Feynman integrals by Mellin--Barnes representation,
Nucl. Phys. Proc. Suppl.}  {\bf 135} (2004) 252
[arXiv:hep-ph/0406052];
\\
G.~Heinrich and V.A.~Smirnov,
{\em Analytical evaluation of dimensionally regularized massive on-shell double
boxes,
Phys. Lett.}  {\bf B 598} (2004) 55
[arXiv:hep-ph/0406053];
\\
Z.~Bern,  L.J.~Dixon, and V.A.~Smirnov,
{\em Iteration of planar amplitudes in maximally supersymmetric Yang--Mills
theory at three loops and beyond,
Phys. Rev.} {\bf D 72} (2005) 085001
[arXiv:hep-th/0505205];
\\
A.G.~Grozin, A.V.~Smirnov and V.A.~Smirnov,
{\em Decoupling of heavy quarks in HQET,
JHEP} {\bf 0611} (2006) 022
[arXiv:hep-ph/0609280];
\\
B.~Jantzen and V.A.~Smirnov,
{\em The two-loop vector form factor in the Sudakov limit,
Eur. Phys. J.} {\bf C 47} (2006) 671
[arXiv:hep-ph/0603133];
\\
J.M.~Drummond, J.~Henn, V.A.~Smirnov and E.~Sokatchev,
{\em Magic identities for conformal four-point integrals,
JHEP} {\bf 0701} (2007) 064
[arXiv:hep-th/0607160].

\bibitem{AnDa}
C.~Anastasiou and A.~Daleo,
{\em Numerical evaluation of loop integrals,
JHEP} {\bf 0610} (2006) 031
[arXiv:hep-ph/0511176].

\bibitem{Czakon}
M.~Czakon,
{\em Automatized analytic continuation of Mellin--Barnes integrals,
Comput. Phys. Commun.}  {\bf 175} (2006) 559
[arXiv:hep-ph/0511200].


\bibitem{SBappl}
E.W.N.~Glover and M.E.~Tejeda-Yeomans,
{\em Progress towards 2 to 2 scattering at two-loops,
Nucl. Phys. Proc. Suppl.}  {\bf 89} (2000) 196
[arXiv:hep-ph/0010031];
\\
C.~Anastasiou, J.B.~Tausk and M.E.~Tejeda-Yeomans,
{\em The on-shell massless planar double box diagram with an irreducible
numerator,
Nucl. Phys. Proc. Suppl.} {\bf 89} (2000) 262
[arXiv:hep-ph/0005328];
\\
C.~Anastasiou, T.~Gehrmann, C.~Oleari, E.~Remiddi and J.B.~Tausk,
{\em The tensor reduction and master integrals of the two-loop massless  crossed
box with light-like legs,
Nucl. Phys.} {\bf B 580} (2000) 577
[arXiv:hep-ph/0003261];
\\
M.~Czakon, J.~Gluza and T.~Riemann,
{\em On the massive two-loop corrections to Bhabha scattering,
Acta Phys. Polon.} {\bf B 36} (2005) 3319
[arXiv:hep-ph/0511187];
\\
T.~Becher and M.~Neubert,
{\em Toward a NNLO calculation of the $\bar{B}\to X_s \gm$
decay rate with  a cut on photon energy. II: Two-loop result for the jet function,
Phys. Lett.}  {\bf B 637} (2006) 251
[arXiv:hep-ph/0603140];
\\
Z.~Bern, M.~Czakon, D.A.~Kosower, R.~Roiban and V.A.~Smirnov,
{\em Two-loop iteration of five-point N = 4 super-Yang--Mills amplitudes,
Phys. Rev. Lett.}  {\bf 97} (2006) 181601
[arXiv:hep-th/0604074];
\\
Z.~Bern, M.~Czakon, L.J.~Dixon, D.A.~Kosower and V.A.~Smirnov,
{\em The four-loop planar amplitude and cusp anomalous dimension in maximally
supersymmetric Yang--Mills theory,
Phys. Rev.} {\bf D 75} (2007) 085010
[arXiv:hep-th/0610248];
\\
T.~Gehrmann, G.~Heinrich, T.~Huber and C.~Studerus,
{\em Master integrals for massless three-loop form factors: One-loop and
two-loop insertions,
Phys. Lett.}  B {\bf 640} (2006) 252
[arXiv:hep-ph/0607185];
G.~Heinrich, T.~Huber and D.~Maitre,
{\em Master integrals for fermionic contributions to massless three-loop form
factors,
Phys. Lett.} {\bf B 662} (2008) 344
[arXiv:0711.3590 [hep-ph]];
\\
A.V.~Smirnov, V.A.~Smirnov and M.~Steinhauser,
{\em Applying Mellin--Barnes technique and Gr\"obner bases to the three-loop
static potential,
PoS} {\bf RADCOR2007} (2007) 024
[arXiv:0805.1871 [hep-ph]];
{\em Evaluating the three-loop static quark potential,
Nucl. Phys. Proc. Suppl.}  {\bf 183} (2008) 308
[arXiv:0807.0365 [hep-ph]];
\\
M.B.~Green, J.G.~Russo and P.~Vanhove,
{\em Modular properties of two-loop maximal supergravity and connections with
string theory,
JHEP} {\bf 0807} (2008) 126
[arXiv:0807.0389 [hep-th]];
\\
M.~Czakon, A.~Mitov and S.~Moch,
{\em Heavy-quark production in massless quark scattering at two loops in QCD,
Phys. Lett.} {\bf B 651} (2007) 147
[arXiv:0705.1975 [hep-ph]];
\\
M.~Czakon,
{\em Tops from light quarks: full mass dependence at two-loops in QCD,
Phys. Lett.} {\bf B 664} (2008) 307
[arXiv:0803.1400 [hep-ph]];
\\
M.~Beneke, T.~Huber and X.Q.~Li,
{\em Two-loop QCD correction to differential semi-leptonic $b \to u$ decays in the
shape-function region},
[arXiv:0810.1230 [hep-ph]];
\\
A.~Pak and A.~Czarnecki,
{\em Heavy-to-heavy quark decays at NNLO},
[arXiv:0808.3509 [hep-ph]];
\\
I.~Bierenbaum, J.~Blumlein, S.~Klein and C.~Schneider,
{\em Two-loop massive operator matrix elements for unpolarized heavy flavor
production to $O(\epsilon)$,
Nucl. Phys.} {\bf B 803} (2008) 1
[arXiv:0803.0273 [hep-ph]];
\\
G.~Somogyi and Z.~Trocsanyi,
{\em A subtraction scheme for computing QCD jet cross sections at NNLO:
integrating the subtraction terms I,
JHEP} {\bf 0808} (2008) 042
[arXiv:0807.0509 [hep-ph]].

\bibitem{CMM}
M.~Czakon, A.~Mitov and S.~Moch,
{\em Heavy-quark production in gluon fusion at two loops in QCD,
Nucl. Phys.} {\bf B 798} (2008) 210
[arXiv:0707.4139 [hep-ph]].

\bibitem{AMBRE}
J.~Gluza, K.~Kajda and T.~Riemann,
{\em AMBRE - a Mathematica package for the construction of Mellin--Barnes
representations for Feynman integrals,
Comput. Phys. Commun.}  {\bf 177} (2007) 879
[arXiv:0704.2423 [hep-ph]].

\bibitem{Gons}
R.J.~Gonsalves,
{\em Dimensionally regularized two loop on-shell quark form-factor,
Phys. Rev.} {\bf D 28} (1983) 1542.

\bibitem{MMV}
S.~Moch, J.A.M.~Vermaseren and A.~Vogt,
{\em The quark form factor at higher orders,
JHEP} {\bf 0508} (2005) 049
[arXiv:hep-ph/0507039].

\bibitem{2lffgenEp}
T.~Gehrmann, T.~Huber and D.~Maitre,
{\em Two-loop quark and gluon form factors in dimensional regularisation,
Phys. Lett.} {\bf B 622} (2005) 295
[arXiv:hep-ph/0507061].

\end{thebibliography}
\end{document}